\UseRawInputEncoding
\pdfoutput=1

\documentclass[twocolumn]{revtex4}
\newlength{\widthfigcolumn}
\newlength{\widthfigpage}
\usepackage{graphicx}
\usepackage[caption=false, subrefformat=parens,labelformat=parens]{subfig}

\usepackage[]{amsmath}
\usepackage[]{algorithm}
\usepackage[noend]{algpseudocode}
\alglanguage{pseudocode}
\usepackage{algorithmicx,eqparbox,array}
\usepackage{mathtools}
\usepackage{booktabs}
\usepackage{amsthm}

\usepackage{placeins}

\algnewcommand{\LongComment}[1]{\hfill// \begin{minipage}[t]{\eqboxwidth{COMMENT\thealgorithm}}#1\strut\end{minipage}}

\begin{document}
	
\title[Mechanical Properties of Silicon Nanowires with Native Oxide Surface State]{Mechanical Properties of Silicon Nanowires with Native Oxide Surface State}

\author{Sina Zare Pakzad}
\affiliation{Department of Mechanical Engineering, Ko\c{c} University, 34450 Sariyer, Istanbul, Turkey}
\email[]{szarepakzad18@ku.edu.tr}

\author{Mohammad Nasr Esfahani}
\affiliation{School of Physics, Engineering and Technology, University of York, York, YO10 5DD, UK}
\email[]{mohammad.nasresfahani@york.ac.uk}

\author{B. Erdem Alaca}
\affiliation{Department of Mechanical Engineering \& n${^2}$STAR-Ko\c{c} University Nanofabrication and Nanocharacterization Center \& Ko\c{c} University Surface Technologies Research Center (KUYTAM), Ko\c{c} University, 34450 Sariyer, Istanbul, Turkey}
\email{ealaca@ku.edu.tr}

\begin{abstract}
Silicon nanowires have attracted considerable interest due to their wide-ranging applications in nanoelectromechanical systems and nanoelectronics. Molecular dynamics simulations are powerful tools for studying the mechanical properties of nanowires. However, these simulations encounter challenges in interpreting the mechanical behavior and brittle to ductile transition of silicon nanowires, primarily due to surface effects such as the assumption of an unreconstructed surface state. This study specifically focuses on the tensile deformation of silicon nanowires with a native oxide layer, considering critical parameters such as cross-sectional shape, length-to-critical dimension ratio, temperature, the presence of nano-voids, and strain rate. By incorporating the native oxide layer, the article aims to provide a more realistic representation of the mechanical behavior for different critical dimensions and crystallographic orientations of silicon nanowires. The findings contribute to the advancement of knowledge regarding size-dependent elastic properties and strength of silicon nanowires.

\end{abstract}
	
\maketitle

\section{Introduction}

Nanowires (NWs) have gained significant attention due to their applications in high-performance devices across various fields, including nanoelectromechanical systems (NEMS) \cite{bachtold2022mesoscopic}, nanoelectronics \cite{li2023sub, zhang2021nanowire}, energy harvesting \cite{nehra20201d}, and electromechanical/biochemical sensors \cite{ alaca2023piezoresistive, akbari2020silicon}. Silicon (Si) NWs have been extensively utilized in various applications, including NEMS for mass spectrometers \cite{liu2022tissue}, gate-all-around transistors \cite{singh2023perspective}, and photodetectors/solar cells \cite{wang2021leaky}. As the size of Si NWs decreases, the surface contribution becomes more significant, leading to a size-dependence in the mechanical behavior of the NWs \cite{nasr2019review}. The investigation of size-dependent mechanical properties in NWs involves both experimental and modeling approaches where both disciplines encounter distinct difficulties \cite{nasr2019review, wang2017mechanical}. Precise study of the scale-dependence in Si NWs through experimental studies poses challenges in sample fabrication, device calibration, high-resolution measurements, NW alignment, modeling methods, and integration with microscale structures \cite{dasgupta201425th, zhu2017mechanics, Elhebeary2017}. Theoretical and computational approaches, such as molecular dynamics (MD) simulations, first-principle methods, and Monte Carlo simulations, are powerful for investigating the mechanical properties of NWs \cite{nasr2019review, momeni2020multiscale}. However, the study of the elastic properties and strength of Si NWs using MD simulations faces inherent complexities and substantial challenges in interpreting their mechanical behavior. \cite{pakzad2021molecular,zare2023nanomechanical, pakzad2023role, nasr2019review}. The challenges primarily arise from the surface effects, which include factors such as the surface state (native oxide vs. unreconstructed surface) and surface properties like surface energy, surface stress, and surface elastic constants \cite{pakzad2021molecular, pakzad2023role, zare2023nanomechanical}. The utilization of the unreconstructed surface condition in MD modeling of the mechanical properties of Si NWs has given rise to discrepancies between computational and experimental findings \cite{yang2022review, nasr2019review, pakzad2023role, zare2023nanomechanical}. Furthermore, comprehending the computational parameters, such as selecting suitable interatomic potentials, determining modeling details like time step and thermodynamic ensembles, as well as defining boundary conditions (B.C.s) and incorporating defects in MD simulations, has become increasingly challenging \cite{nasr2019review, pakzad2023role, pakzad2021molecular, zare2023nanomechanical}. In this context, the presence of native oxide, specifically amorphous silicon dioxide, as a vital component of the Si NW surface has a significant impact on the NW structure by introducing defects or other interfaces \cite{nasr2022effect, pakzad2023role, zare2023nanomechanical}. Recent MD studies demonstrate that the modulus of elasticity of Si NWs can be reduced by up to 40\% due to the native oxide surface condition compared to the unreconstructed surface state \cite{pakzad2023role}. Additionally, a similar reduction of up to 20\% is estimated in the ultimate strength \cite{pakzad2023role}. Other structural aspects such as finite deformation and intrinsic stresses have to be modeled for mechanical consistency and relevance with respect to experimental conditions. A recent study has brought attention to the tensile intrinsic stresses in Si NWs that can reach up to 1.2 GPa when native oxide forms under ambient conditions \cite{nasr2022effect}. The clear reduction observed in the mechanical properties of Si NWs highlights the need for further detailed examination, particularly considering the minuscule variations in the thickness of the native oxide layer and the cross-sectional shape \cite{zare2021new, pakzad2023role, pakzad2021molecular, zare2023nanomechanical}.

Despite its significant importance, there are still several unresolved issues concerning the mechanical behavior of Si NWs \cite{minor2005room}. One such topic is brittle-to-ductile transition (BDT), which lacks detailed \emph{in-situ} thermomechanical testing and computational modeling at the nanoscale \cite{cheng2019situ, kang2007brittle}. There is growing evidence indicating that ductile behavior is observed in reduced-size Si NWs at much lower temperatures \cite{han2007low, cheng2019situ}. For instance, BDT is observed for Si NWs under tension at a critical dimension of 20 nm and a temperature as low as $90 ^\circ$C, whereas in bulk Si, BDT typically occurs at much higher temperatures, such as $540^\circ$C \cite{cheng2019situ}. Limited literature exists regarding bending, where BDT has also been identified as a scale-dependent and non-intrinsic property \cite{zheng2009atomic}. In this respect, a recent bending study at $400^\circ$C reported activation of $<110>$ \{111\} dislocations at a much larger scale (2-5 mm) with multiple dislocation nucleation sites appearing on the high-stress surface \cite{Elhebeary2020time}. Numerical modeling studies have explored the BDT phenomenon in Si NWs at the atomic level, yielding varying results. Some studies investigated the fracture behavior of Si NWs through MD simulations \cite{kang2010size, liu2012large, guenole2011deformation, el2015onset, godet2015surface, jeng2005effects, xu2020molecular, kuo2014mechanical}. For instance, MD simulations conducted on Si NWs with diameters ranging from 2-8 nm found a clear size and temperature dependency, with only NWs below 4 nm exhibiting ductility at low temperatures \cite{kang2010size}. Another study demonstrated through MD simulations that the competition between surface facet effect and axial orientation controls Si NW BDT, leading to different failure modes like dislocation emission, crystal-amorphous transition, or cleavage \cite{liu2012large}. The existence of both brittle and ductile behavior at 10 K indicates a potential shift in the transition size at low temperatures, as indicated by experimental trends \cite{liu2012large}. Regarding this matter, a constraint in BDT studies conducted using the MD method lies in their assumption of an unreconstructed surface state. This limitation makes the comparison between computational and experimental results difficult, as the experimental conditions may involve different surface states \cite{nasr2022effect}. Moreover, numerous MD studies have shown significant variation in the predicted fracture mechanisms. Some studies have identified dislocation slip as the dominant mechanism \cite{godet2009evidence, yang2009shape}, while others have suggested that the initiation of amorphization is energetically favorable around the crack tip \cite{han2007low, huang2009mechanics, buehler2007threshold}. To address this, failure probability formulations that integrate the ductile/brittle fracture mechanism into MD studies have been developed \cite{xu2019molecular, kang2007brittle, pakzad2023role}.

Moreover, the examination of mechanical properties of Si NWs as a function of temperature using MD method has also been limited to the unreconstructed surface conditions \cite{kang2007brittle, xu2019molecular, Jing2009, park2005molecular, Jing2009b}. Another fundamental difference between experiments and atomistic simulations dedicated to the study of mechanical properties is the time scale difference which leads to the use of different strain rates that are often several orders of magnitude higher than real experiments \cite{kang2010size, xu2019molecular, pakzad2023role}. In this respect, studies dedicated to analysis of strain rate effect of elastic properties and strength of Si NWs are common. A recent study exhibited deviations up to 10 \% and 15 \% for the modulus of elasticity and fracture strength of Si NWs with native oxide surface conditions \cite{pakzad2023role}. With the exception of a single study \cite{pakzad2023role}, other studies investigating the strain rate effect on Si NWs using the MD method have also benefited from the unreconstructed surface assumption. This combination, along with the selection of interatomic potential and critical dimension of Si NWs, has led to the observation of diverse effects \cite{kang2010size, xu2019molecular}. Moreover, the geometrical effects of Si NWs such as length to critical dimension ratio \cite{pakzad2023role} and the shape of cross-section \cite{Jing2009b, Liu2015, vogl2021effect} has been the topic of MD studies through tensile and bending tests reporting significant effects on mechanical behavior of Si NWs. The presence of flaws and defects can alter the properties of Si NWs and pose a critical issue in terms of atomic structure arrangement leading to changes in elastic properties and strength of NWs \cite{samykano2021progress, Elhebeary2020time}. To comprehend the impact of such imperfections on the mechanical properties of NWs, extensive investigations are conducted on various forms of deficiencies, including point defects, vacancy clusters, defective surfaces, surface steps, and nano-voids \cite{huang2011coupled, Liu2015, katakam2022tensile, cheng2015large}.

Motivated by the discrepancies observed between computational and experimental findings on the mechanical properties of Si NWs, this study focuses on investigating the tensile deformation of Si NWs. To tackle this issue, the present MD study on Si NWs with a native oxide surface state incorporates several critical parameters mentioned earlier. The subsequent sections of this article provide detailed information regarding the simulation methodology used to model the tensile response of Si NWs with native oxide surface state. The tensile simulations aim to investigate the influence of various factors, including cross-sectional shape, length to critical dimension ratio, temperature, the presence of nano-voids, and strain rate on the mechanical behavior of Si NWs with different sizes and crystalline orientations. The objective of this article is to bridge the existing discrepancies between atomistic studies and experimental investigations by considering the more realistic surface state of native oxide in Si NWs.

\section{Materials and Methods}

\begin{figure*}[ht]
	\centering
	\includegraphics[width=480pt]{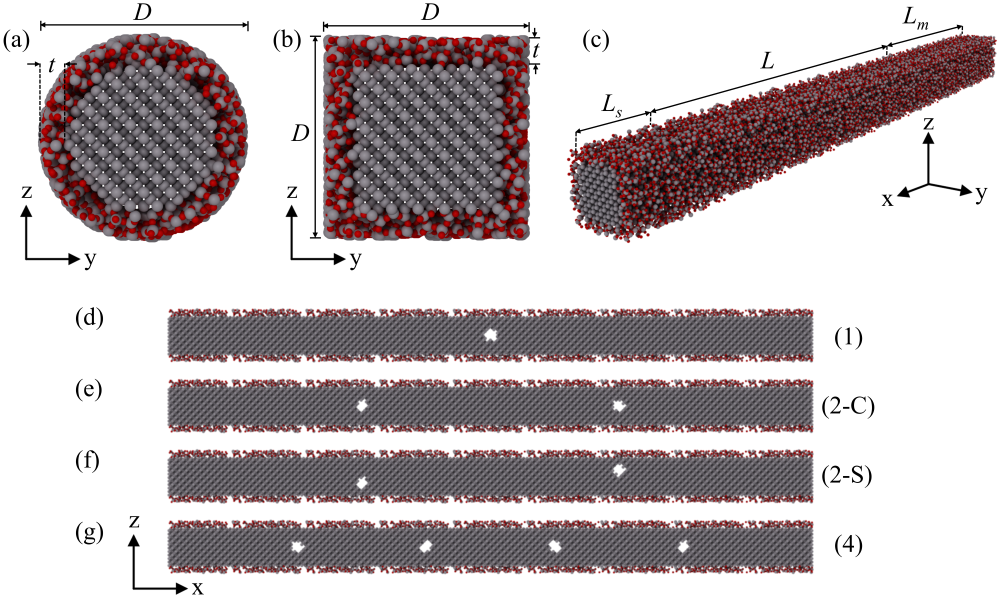}
	\caption{\label{fig:fig1} Atomic configurations of aSiO$_2$-Si NWs with different cross-sectional shapes. (a) Circular cross-section and (b) square cross-section, where the critical dimension, $D$, and native oxide thickness, $t$, are indicated. (c) Representative view of an aSiO$_2$-Si NW, illustrating the lengths: NW length ($L$), support region length ($L_s$), and movable region length ($L_m$). Si and O atoms are represented by grey and red spheres, respectively. Side-views of aSiO$_2$-Si NWs with various nano-void configurations: (d) single (1), (e) two centered (2-C), (f) two surface (2-S), and (g) four (4) nano-voids distributed along the NWs.}
\end{figure*}

Tensile simulations are performed on Si NWs with a native oxide surface state using the MD method \cite{thompson2022lammps}. Figure~\ref{fig:fig1} (a) and Figure~\ref{fig:fig1} (b) depict the initial atomic configurations of Si NWs with two different cross-sectional shapes: circular and square, respectively. The critical dimension, denoted as $D$, ranges from 4 nm to 8 nm, representing the diameter for circular NWs and the width/height for square NWs. The native oxide layer, characterized by an amorphous structure, is referred to as aSiO$_2$, and Si NWs with native oxide are denoted as aSiO$_2$-Si NWs throughout the remainder of this work. For the aSiO$_2$-Si NWs in this study, a native oxide layer thickness ($t$) of 0.5 nm is defined. The length ($L$) of the aSiO$_2$-Si NWs remains constant with a length-to-critical dimension ($L/D$) aspect ratio (AR) of 10, except for the tests conducted to examine the AR effect. Figure~\ref{fig:fig1} (c) illustrates a Si NW with its length denoted as $L$, where the support and movable boundaries are labeled as $L_s$ and $L_m$, respectively. Specifically, the boundary lengths $L_s$ and $L_m$ are set to one-eighth of the NW length ($L$). By maintaining an AR of 10 across different critical dimensions of aSiO$_2$-Si NWs, any potential length-related effects are effectively eliminated \cite{pakzad2023role}. MD simulations are conducted on aSiO$_2$-Si NWs oriented along $<100>$, $<110>$, $<111>$, and $<112>$ crystal orientations. Non-periodic B.C.s are applied along all directions ($x$, $y$, and $z$) to account for the finite size of the system. In order to ensure consistency in cross-section designs and to thoroughly examine the effect of surface state, it is common to observe rectangular or square geometries in top-down fabricated Si NWs with crystal orientations $<100>$ and $<110>$ \cite{nasr2019review}. Conversely, circular cross-sections are frequently encountered in bottom-up fabricated Si NWs with crystal orientations $<111>$ and $<112>$ \cite{nasr2019review}. Therefore, this study models aSiO$_2$-Si NWs by considering both cross-sectional shapes and four crystal orientations of Si.

Atoms are initially assigned an initial velocity with a Gaussian distribution at a finite temperature. Afterward, an energy minimization is performed using the conjugate gradient method. Following this, the NW is relaxed at a constant temperature for 50 ps, employing a time step of 1 fs in a canonical NVT ensemble. During the relaxation step, the support and movable regions connected to the NW are held in a fixed configuration. In the subsequent step, a tensile test is performed by fixing one side of the NW ($L_s$) while allowing the other end ($L_m$) to move. To prevent the occurrence of shock waves due to high loading rates, a linearly increasing constant velocity is applied along the longitudinal direction ($x$-direction). This velocity starts from zero at the support and gradually reaches its maximum value at the movable boundary. In this regard, a velocity of 1.0 \AA/ps is utilized at a constant temperature, corresponding to a strain rate of $\dot{\epsilon} \approx 1.0 \times 10^{9}$ s$^{-1}$. This strain rate is considered appropriate for conducting tensile tests on NWs using MD simulations, as supported by previous studies \cite{kang2007brittle, pakzad2023role, pakzad2021molecular}. In this work, the modified Stillinger-Weber (m-SW) potential \cite{ganster2010atomistic}, which has shown promising results in modeling Si NWs with native oxide, is employed. The specific expressions and coefficients for the m-SW potential can be found in the associated reference \cite{ganster2010atomistic}. Within this study, various parameters such as critical dimension, crystallographic orientation, AR, strain rate, and temperature are examined for aSiO$_2$-Si NWs. Additionally, the impact of nano-voids, including their number and placement, on the mechanical properties of the NWs is investigated. Figures~\ref{fig:fig1} (d-g) display atomic snapshots of aSiO$_2$-Si NWs with different nano-void configurations: single (1), two centered (2-C), two surface (2-S), and four (4) nano-voids placed along the NWs. The diameter of the spherical voids is regarded as $1/4^{th}$ of $D$ for NWs of different sizes. The surface-based nano-voids (2-S) are positioned near the crystalline Si core and the native oxide surface layer, allowing for potential comparisons with the centered (2-C) nano-voids given in Figure~\ref{fig:fig1} (f) and Figure~\ref{fig:fig1} (e), respectively.

The stress state, $\pi_{ij}$, of the associated aSiO$_2$-Si NWs is determined using the virial theorem \cite{zimmerman2004calculation}, as given in Equation~\ref{eqn1}. Here, $\Omega_0$ represents the atomic volume in the undeformed system, with $N$ being the total number of atoms. The atomic volumes for Si and aSiO$_2$ are obtained through the relaxation of the corresponding structures, and further details can be found in Ref. \cite{pakzad2021molecular}. The atomic distances between atoms $\alpha$ and $\beta$ are denoted as $r^{\alpha\beta}$. The position of atom $\alpha$ along the $j$ direction is represented by $v_{j}^{\alpha}$, and the difference in positions between atoms $\alpha$ and $\beta$ is given as $v_{j}^{\alpha\beta} = v_{j}^{\alpha}-v_{j}^{\beta}$. The interatomic potential is represented by $V$.

\begin{equation}
	\label{eqn1}
	\pi_{ij} = \frac{1}{2\Omega_{0}}\bigg[\sum_{\alpha=1}^{N}\sum_{\beta \neq \alpha}^{N} \frac{1}{r^{\alpha\beta}} \frac{\partial{V}(r^{\alpha\beta})}{\partial r} (v_{i}^{\alpha\beta}v_{j}^{\alpha\beta} )\bigg]
\end{equation}

\section{Results}
\label{sec:results}

The section begins by establishing the stress-strain curves for different critical dimension and cross-sectional shapes of Si NWs for four crystallographic orientation of core-Si given in Section~\ref{sec:3.1}. Furthermore, the elastic properties and strength of Si NWs are studied as a function of AR (Section~\ref{sec:3.2}), temperature (Section~\ref{sec:3.3}), nano-void effects (Section~\ref{sec:3.4}), and strain rate (Section~\ref{sec:3.5}).

\subsection{Cross-sectional Shape Effects}
\label{sec:3.1}

\begin{figure}[ht]
	\centering
	\includegraphics[width=0.45\textwidth]{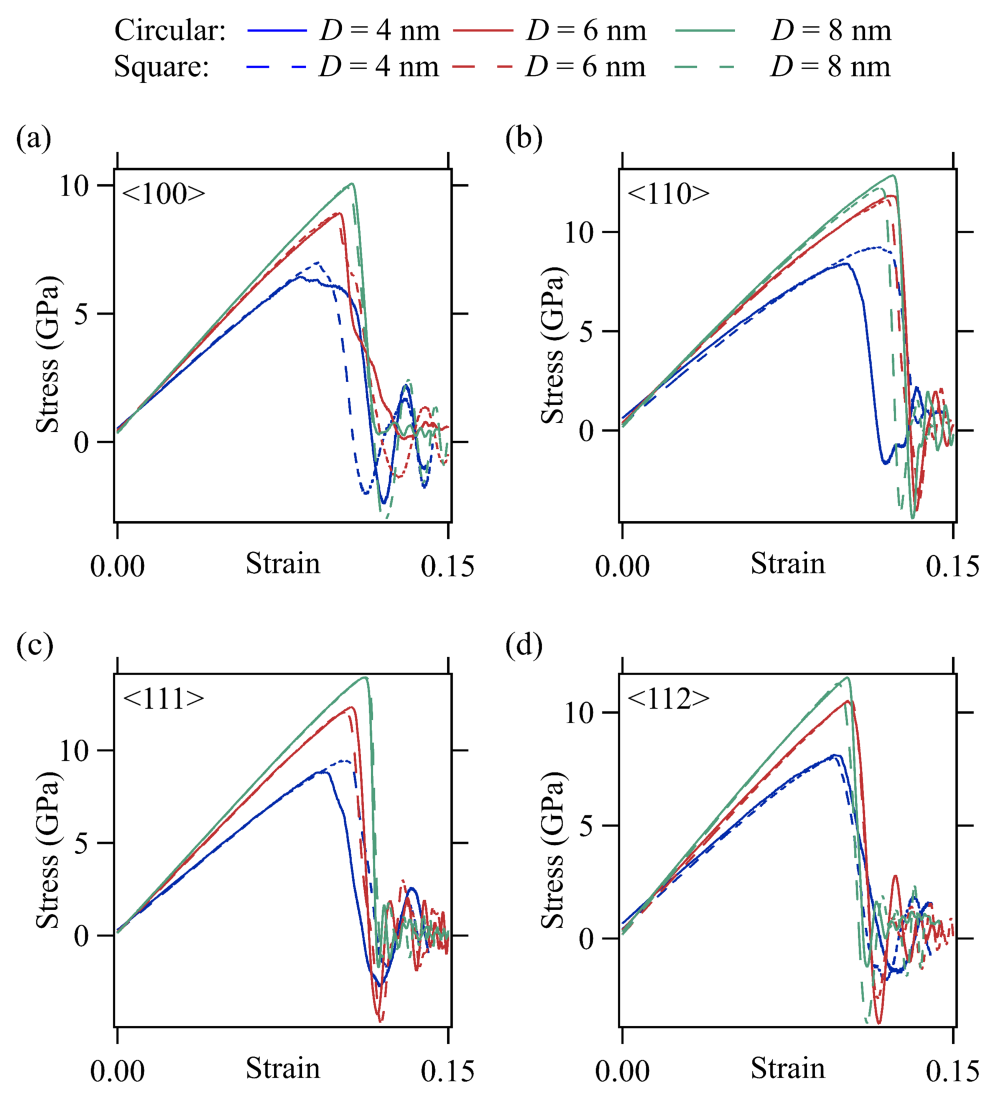}
	\caption{\label{fig:fig2} Stress-strain curves of aSiO$_2$-Si NWs of different critical dimensions with an AR of 10. The curves correspond to NWs oriented along (a) $<100>$, (b) $<110>$, (c) $<111>$, and (d) $<112>$ crystal orientations. Solid lines represent NWs with circular cross-sections, while dashed lines correspond to NWs with square cross-sections.}
\end{figure}

Figure~\ref{fig:fig2} displays the stress-strain curves for aSiO$_2$-Si NWs with circular and square cross-sections and critical dimensions of 4 nm, 6 nm, and 8 nm. The NWs are oriented along $<100>$, $<110>$, $<111>$, and $<112>$ crystal orientations given in Figure~\ref{fig:fig2} (a), (b), (c), and (d), respectively. The design of the aSiO$_2$-Si NWs includes an AR of 10 and a native oxide thickness of 0.5 nm, enabling a comprehensive examination of the combined effects of crystal orientation and cross-sectional shape. In the elastic regime, characterized by small strains, the stress-strain relationship is linear, allowing for the direct prediction of the modulus of elasticity ($E$). As the stress increases, the relationship becomes non-linear until it reaches the ultimate strength ($S$). Subsequently, the stress starts to decrease due to necking and fracture of the NWs. The ultimate strength is determined as the maximum stress achieved during the tensile deformation. The stress-strain curves presented in Figure~\ref{fig:fig2} are utilized to measure the modulus of elasticity and ultimate strength for various crystal orientations and cross-sectional shapes. The results are summarized in Table~\ref{tab:table1}. The findings demonstrate that as the critical dimension increases, the modulus of elasticity and ultimate strength of the NWs also increase. This comparison was made while maintaining the AR and $t$ constant. Specifically, the modulus of elasticity shows a significant change, ranging from 25\% for $<100>$ and $<110>$ crystal orientations to 30\% for $<111>$ and $<112>$ crystal orientations when comparing $D$ = 4 nm to $D$ = 8 nm. Moreover, there is an observed increase of up to 40\% in the ultimate strength of aSiO$_2$-Si NWs with the increment in $D$. The influence of cross-sectional shape on the predicted properties,  summarized in Table~\ref{tab:table1}, results in deviations of up to 2.5 GPa and 1 GPa for the modulus of elasticity and ultimate strength estimations, respectively. The deviation between the shape-dependent modulus of elasticity and ultimate strength of aSiO$_2$-Si NWs becomes more significant as the critical dimension decreases. The results presented in Section~\ref{sec:3.1} are obtained at a temperature of 1 K and a strain rate of $\dot{\epsilon} \approx 1.0 \times 10^{9}$ s$^{-1}$, while maintaining a constant AR of 10. Further details regarding different ARs for aSiO$_2$-Si NWs are discussed in Section \ref{sec:3.2}. The subsequent discussions on the effects of temperature, nano-void, and strain rate can be found in Section \ref{sec:3.3}, Section \ref{sec:3.4} and Section \ref{sec:3.5}, respectively.

\begin{table*}\centering
	\caption{\label{tab:table1} The table presents the crystallographic orientations and dimensions of aSiO$_2$-Si NWs with circular and square cross-sections, along with their corresponding modulus of elasticity ($E$) and ultimate strength ($S$).}
\begin{ruledtabular}
\begin{tabular}{cccccccccc}
	NW Orientation & \multicolumn{3}{c}{Si Orientation} & $D$ (nm) &  $L$ (nm) & \multicolumn{2}{c}{$E$ (GPa)} &  \multicolumn{2}{c}{$S$ (GPa)}  \\
	\hline\addlinespace[2pt]
	& $x$ & $y$ & $z$ && & Circular & Square	& Circular & Square\\
	\hline\addlinespace[2pt]
	$<100>$ 		 &  [100]  & [010] & [001] & 			4 & 40 & 75.5 & 77.4  & 6.5 & 7.0 \\
	&&&& 													6 & 60 & 90.8 & 90.9  & 8.9 & 8.9 \\
	&&&&													8 & 80 & 95.5 & 95.8  & 10.1 &  10.0 \\
	\hline\addlinespace[2pt]
	$<110>$      	& [110] & [001] &   [$\bar{1}$10]  & 	4 & 40 & 93.3 & 95.7  & 8.4 & 9.3 \\
	&&&& 													6 & 60 & 112.6 & 112.9  & 11.9 & 11.6 \\
	&&&&													8 & 80 & 117.9 & 118.8  & 12.9 &  12.2 \\
	\hline\addlinespace[2pt] 
	$<111>$  		 & [111] & [1$\bar{1}$0]  &  [$\bar{1}$$\bar{1}$2]  &  	4 & 40 & 99.1 & 99.7  & 8.8 & 9.5  \\
	&&&& 																	6 & 60 & 122.2 & 122.0  & 12.3 & 12.1 \\
	&&&&																	8 & 80 & 129.1 & 129.8  & 13.9 &  13.9 \\
	\hline\addlinespace[2pt]
	$<112>$   	 & [112]	  & [11$\bar{1}$] & [1$\bar{1}$0] &  			4 & 40 & 88.8 & 90.2  & 8.2 & 8.0 \\
	&&&& 																	6 & 60 & 107.9 & 108.0  & 10.5 & 10.6 \\
	&&&&																	8 & 80 & 118.4 & 119.0  & 11.6 & 11.3  \\	
\end{tabular}
\end{ruledtabular}
\end{table*}

\subsection{Length to Critical Dimension Effect}
\label{sec:3.2}

\begin{figure}[ht]
	\centering
	\includegraphics[width=0.5\textwidth]{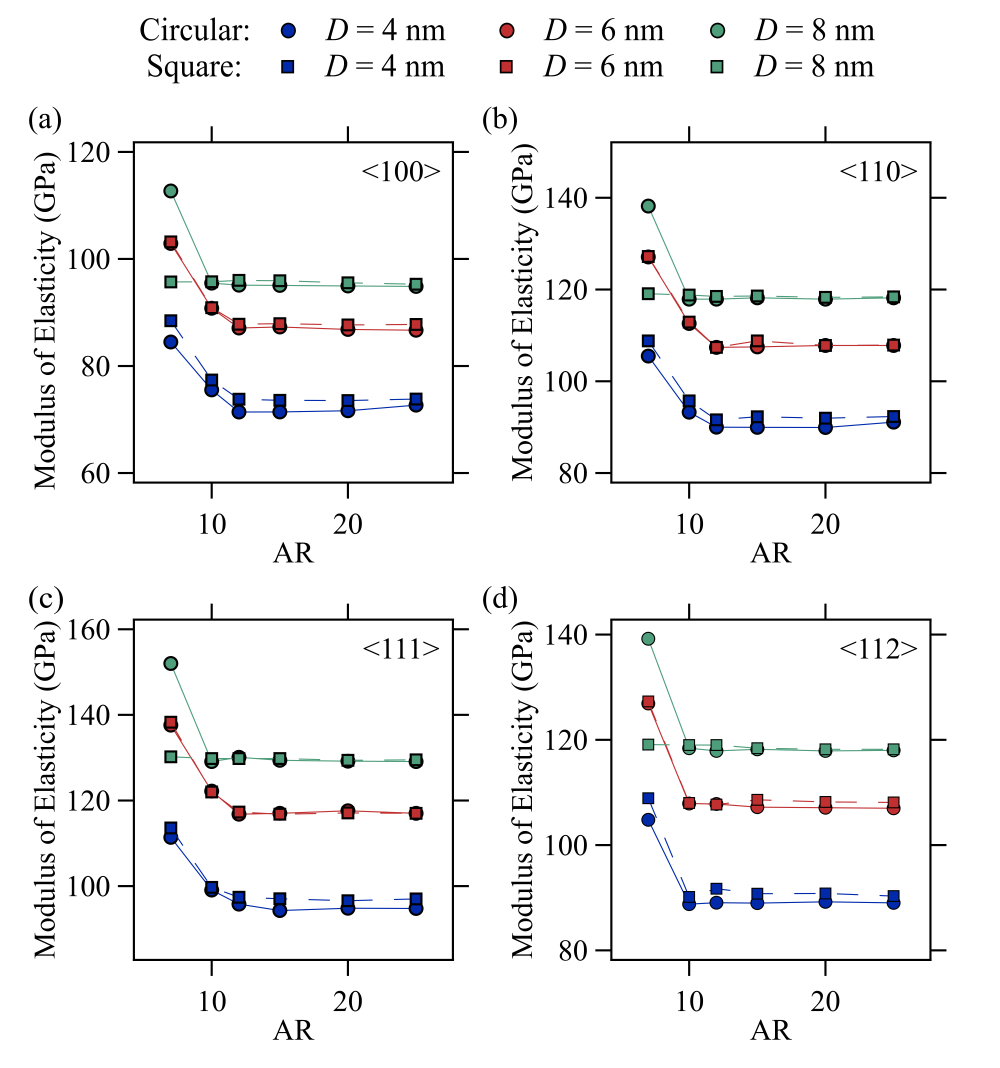}
	\caption{\label{fig:fig3} The modulus of elasticity as a function of AR for aSiO$_2$-Si NWs oriented along (a) $<100>$, (b) $<110>$, (c) $<111>$, and (d) $<112>$ crystal orientations.}
\end{figure}

To further investigate the size-dependent elastic properties and strength of aSiO$_2$-Si NWs, the modulus of elasticity and ultimate strength of NWs with varying ARs are examined. Tensile tests are performed on aSiO$_2$-Si NWs with similar $D$ and crystal orientations as described in Section~\ref{sec:3.1}, with AR values ranging from 7 to 25. Figure~\ref{fig:fig3} illustrates the relationship between the modulus of elasticity and the AR for aSiO$_2$-Si NWs oriented along crystal orientations $<100>$, $<110>$, $<111>$, and $<112>$, as shown in Figure~\ref{fig:fig3} (a), (b), (c), and (d), respectively. The results indicate that there is a negligible size effect on the modulus of elasticity of aSiO$_2$-Si NWs across all crystal orientations when the AR exceeds 12. Additionally, the findings demonstrate that NWs with smaller AR exhibit higher modulus of elasticity, and this trend is consistent across all crystal orientations for NWs with $D$ of 4 nm and 6 nm. For NWs with $D$ = 4 nm, the modulus of elasticity predictions show more variation depending on the shape of the cross-section, whereas for NWs with larger $D$, the shape effect becomes insignificant for AR values greater than 10. Another observation regarding the shape effect is that NWs with a square cross-section ($D$ = 8 nm) at an AR of 7 exhibit a modulus of elasticity similar to higher ARs, while NWs with a circular cross-section demonstrate a higher modulus of elasticity. This could be attributed to the influence of shape effects and surface stresses when modeling NWs under free surface conditions and non-periodic B.C.s. Comparing these findings with previous literature, MD studies on the bending response of Si NWs with unreconstructed surface state indicate a slight decrease in the modulus of elasticity as the AR of the Si NWs increases from 10 to 25 \cite{nasr2020influence}. Another recent MD study comparing the tensile properties of unreconstructed Si NWs and Si NWs with a native oxide revealed an insignificant effect of AR ranging from 7 to 40 when NWs were assumed as part of an infinite NW with periodicity applied along the NW dimension during modeling \cite{pakzad2023role}. However, the present study addresses the length factor by applying non-periodic B.C. along the NW, thereby simulating the length and subsequent AR effect under fixed B.C. When comparing the estimations of the modulus of elasticity with different ARs to experimental findings, the results show reliable agreement, with increased AR accurately reflecting experimental Si NWs, and the estimated values of $E$ falling within the range of prior estimations for different orientations of Si NWs \cite{nasr2019review}.

\begin{figure}[ht]
	\centering
	\includegraphics[width=0.5\textwidth]{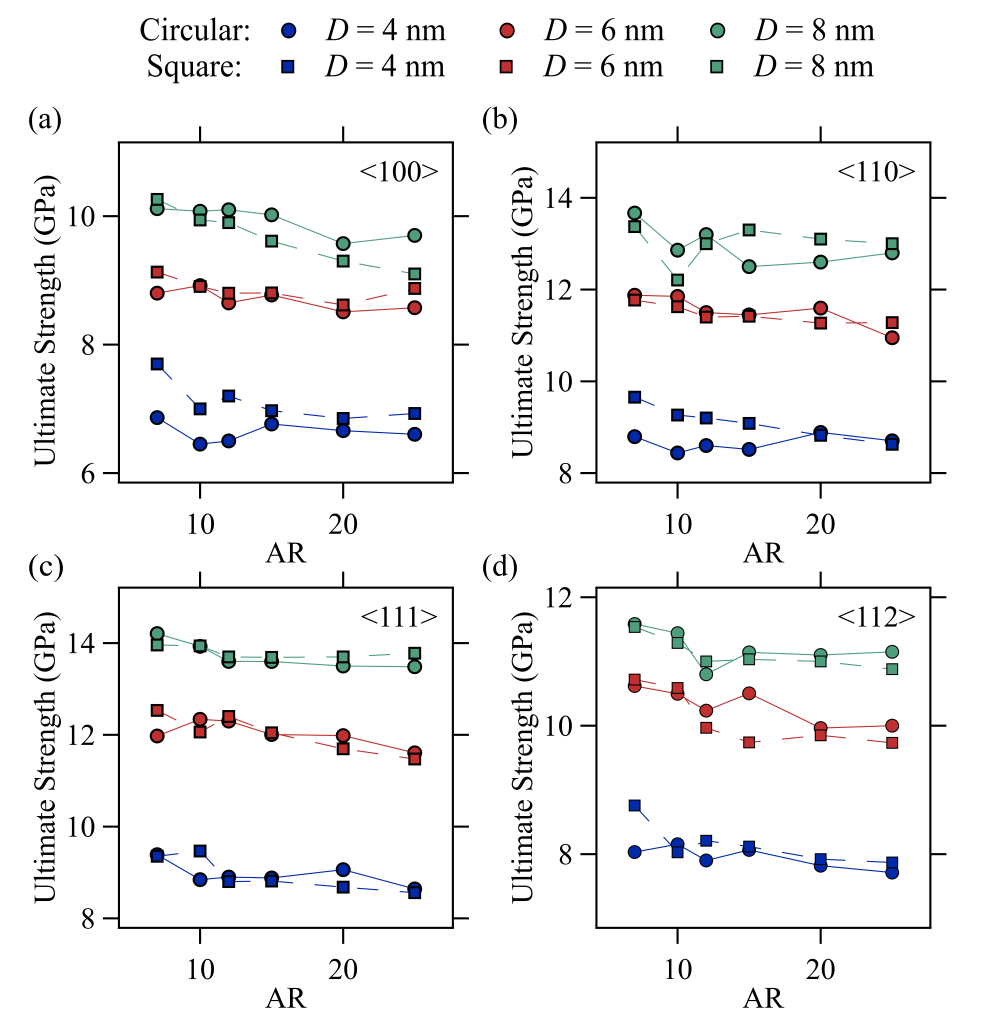}
	\caption{\label{fig:fig4} The ultimate strength as a function of AR for aSiO$_2$-Si NWs oriented along (a) $<100>$, (b) $<110>$, (c) $<111>$, and (d) $<112>$ crystal orientations.}
\end{figure}

Figure~\ref{fig:fig4} presents the correlation between the ultimate strength and the AR for aSiO$_2$-Si NWs oriented along crystal orientations $<100>$, $<110>$, $<111>$, and $<112>$, as shown in Figure~\ref{fig:fig4} (a), (b), (c), and (d), respectively. The results demonstrate slight variations in the ultimate strength for different ARs of aSiO$_2$-Si NWs, influenced by factors such as cross-sectional shape and crystal orientation effects. Moreover, NWs with larger critical dimensions exhibit higher ultimate strength, ranging from 8-9 GPa for $D$ of 4 nm to 10-14 GPa for $D$ of 8 nm. The cross-sectional shape effect on the ultimate strength predictions is influenced by the combined effects of crystal orientation, $D$, and AR of the NWs. Previous works on partially oxidized Si NWs have reported an ultimate strength of approximately 10 GPa, which aligns with the findings depicted in Figure~\ref{fig:fig4} \cite{xu2020molecular}. Another recent study on the tensile properties of Si NWs with native oxide has demonstrated a negligible AR effect when NWs are considered as part of an infinite NW \cite{pakzad2023role}. MD studies on unreconstructed Si NWs have reported ultimate strength ranging from 3 GPa to 15 GPa \cite{kang2007brittle}. It is common to observe a wide range of predictions for the ultimate strength of Si NWs, particularly in experimental studies, where values up to 20 GPa have been reported in the literature \cite{nasr2019review}. The findings presented here indicate that the assumption of a native oxide surface state and the accuracy of the m-SW potential contribute to good agreement with experimental results in estimating the strength of Si NWs.

\subsection{Temperature Effect}
\label{sec:3.3}

\begin{figure}[ht]
	\centering
	\includegraphics[width=0.5\textwidth]{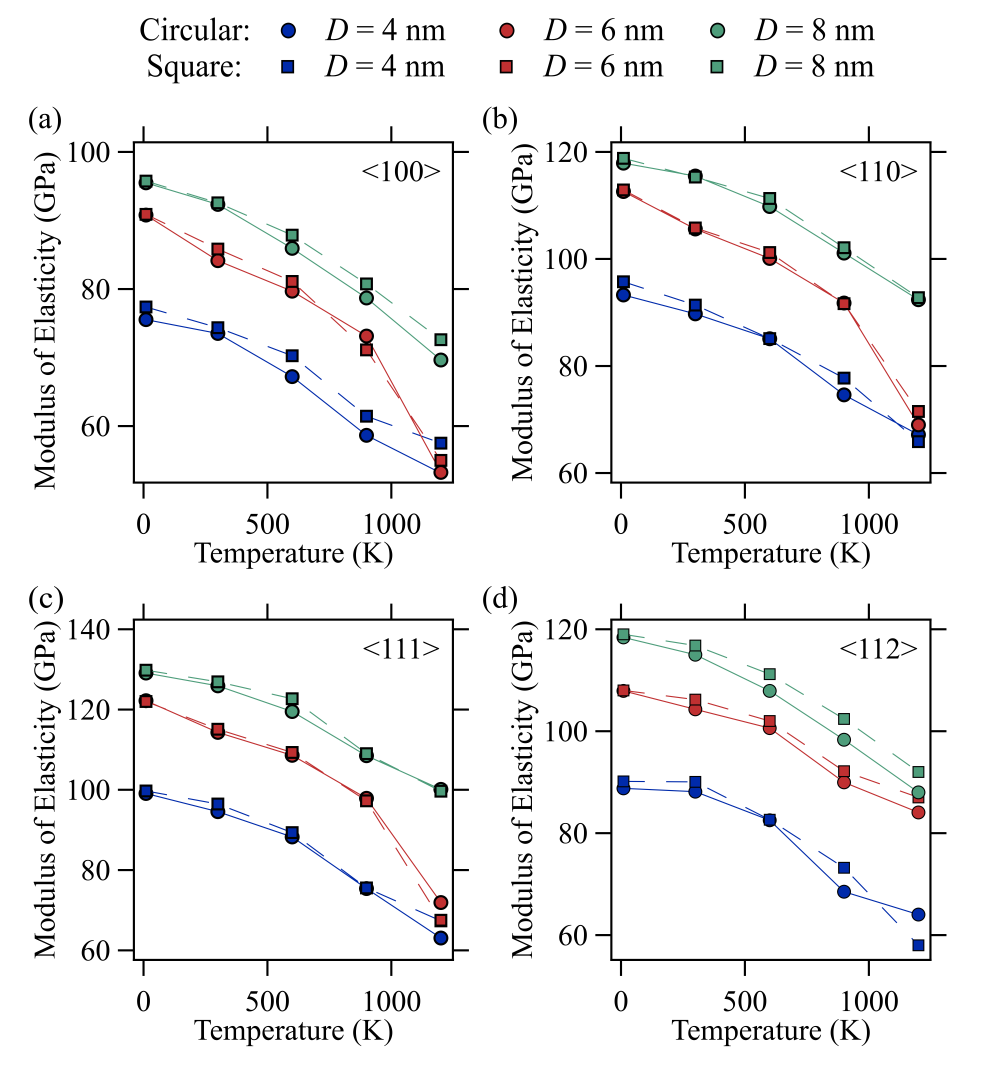}
	\caption{\label{fig:fig5} The modulus of elasticity as a function of temperature for aSiO$_2$-Si NWs oriented along (a) $<100>$, (b) $<110>$, (c) $<111>$, and (d) $<112>$ crystal orientations.}
\end{figure}

\begin{figure}[ht]
	\centering
	\includegraphics[width=0.5\textwidth]{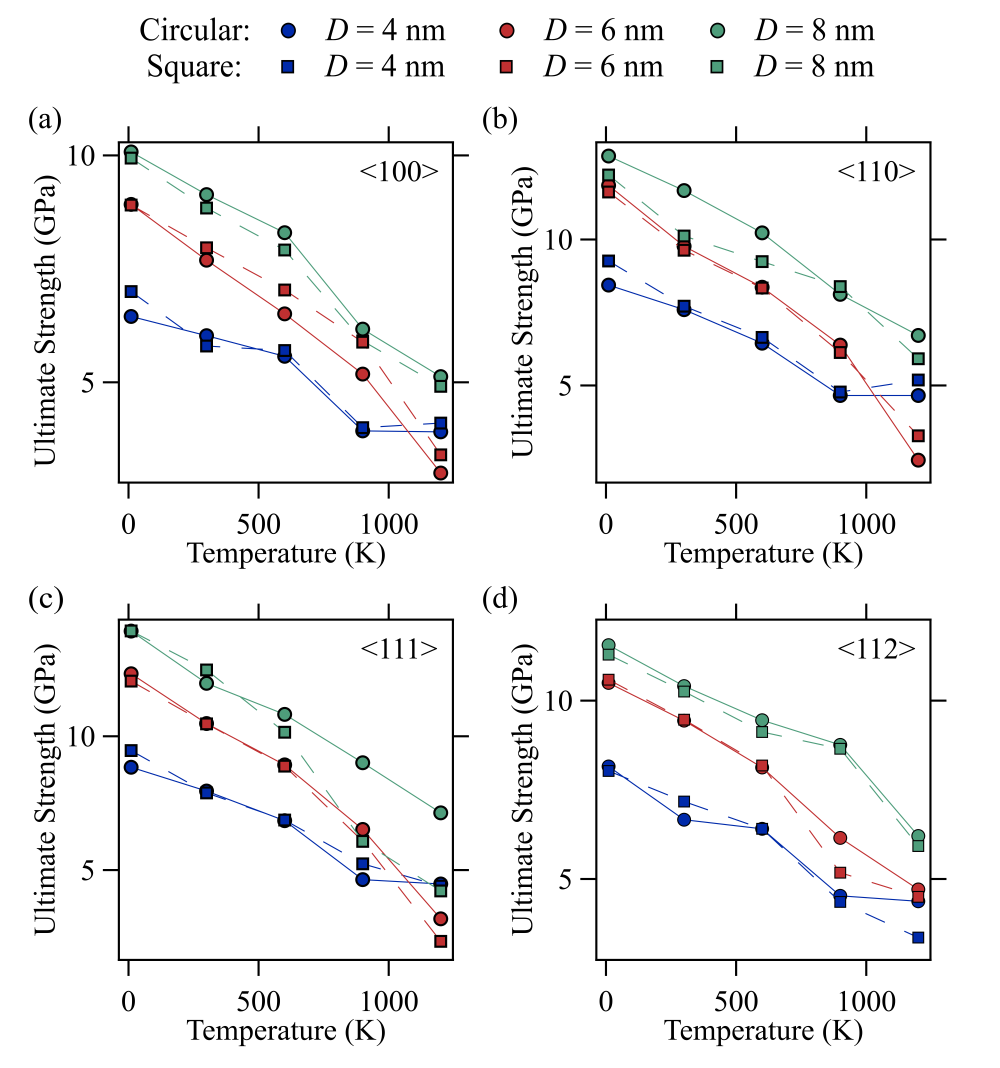}
	\caption{\label{fig:fig6} The ultimate strength as a function of temperature for aSiO$_2$-Si NWs oriented along (a) $<100>$, (b) $<110>$, (c) $<111>$, and (d) $<112>$ crystal orientations.}
\end{figure}

This section focuses on the predictions of the modulus of elasticity and ultimate strength for aSiO$_2$-Si NWs with a particular emphasis on examining their temperature dependence. In the preceding sections of the paper, the tensile simulations for aSiO$_2$-Si NWs were limited to 1 K. However, this section extends the temperature range from 1 K up to 1200 K to explore the temperature-dependent behavior of the NWs. The resulting predictions for the modulus of elasticity and ultimate strength are illustrated in Figure~\ref{fig:fig5} and Figure~\ref{fig:fig6}, respectively. Figure~\ref{fig:fig5} illustrates the modulus of elasticity for aSiO$_2$-Si NWs as a function of temperature oriented along crystal orientations $<100>$, $<110>$, $<111>$, and $<112>$, as depicted in Figure~\ref{fig:fig5} (a), (b), (c), and (d), respectively. Regarding the modulus of elasticity predictions, a significant decrease of approximately 20-50 GPa is observed as the temperature increases for different crystal orientations. Specifically, the temperature rise from 1 K to 1200 K leads to a reduction in the modulus of elasticity from 80-120 GPa at 1 K to 50-100 GPa. To the best of our knowledge, no prior study has reported the temperature-dependent modulus of elasticity for Si NWs with a native oxide surface state. However, comparing our findings with previous MD studies on unreconstructed Si NWs tested under tensile \cite{kang2010size, xu2019molecular} and buckling \cite{Jing2009} conditions, similar deviations in the modulus of elasticity can be observed, resulting in reductions due to temperature increases. The reduction in the modulus of elasticity for unreconstructed Si NWs tested at 100 K and 1000 K is estimated to be up to 100 GPa. In this regard, aSiO$_2$-Si NWs exhibit a lower temperature dependence compared to unreconstructed Si NWs \cite{xu2019molecular}.

Figure~\ref{fig:fig6} shows the ultimate strength estimations as a function of temperature for NWs of different crystal orientations. The results are depicted for aSiO$_2$-Si NWs oriented along crystal orientations $<100>$, $<110>$, $<111>$, and $<112>$, as shown in Figure~\ref{fig:fig6} (a), (b), (c), and (d), respectively. The results indicate a decrease in the ultimate strength as the temperature increases. The magnitude of this reduction varies depending on factors such as critical dimension, cross-sectional shape, and crystal orientation. Specifically, the ultimate strength decreases by up to 12 GPa when the temperature is increased from 1 K to 1200 K. For different $D$ and crystal orientations, the ultimate strength, which ranges between 6-14 GPa at 1 K, decreases to 1-4 GPa at 1200 K. Deviation of up to 2 GPa is observed in the ultimate strength, emphasizing the significant impact of the cross-sectional shape on the strength of aSiO$_2$-Si NWs. Prior MD studies on the buckling behavior of unreconstructed Si NWs have demonstrated a decreasing trend in the critical load with increasing temperature \cite{Jing2009}. Similarly, another MD study investigating the temperature effect on unreconstructed Si NWs under tensile testing reported a reduction in ultimate strength from 12-14 GPa to 4-5 GPa as the temperature increased from 10 K to 1200 K \cite{kang2010size}. Furthermore, an MD study addressing the temperature impact on the ultimate strength of unreconstructed Si NWs found a reduction from 15 GPa to 5-10 GPa for different combinations of critical dimensions and interatomic potentials as the temperature increases from 100 K to 1000 K \cite{xu2019molecular}. In this respect, the observed trend of reduction in ultimate strength for Si NWs with a native oxide surface state is consistent with that of unreconstructed Si NWs, exhibiting reductions that depend on the critical dimensions, crystal orientation, and cross-sectional shape of NWs.

\subsection{Nano-void Effect}
\label{sec:3.4}

\begin{figure}[ht]
	\centering
	\includegraphics[width=0.5\textwidth]{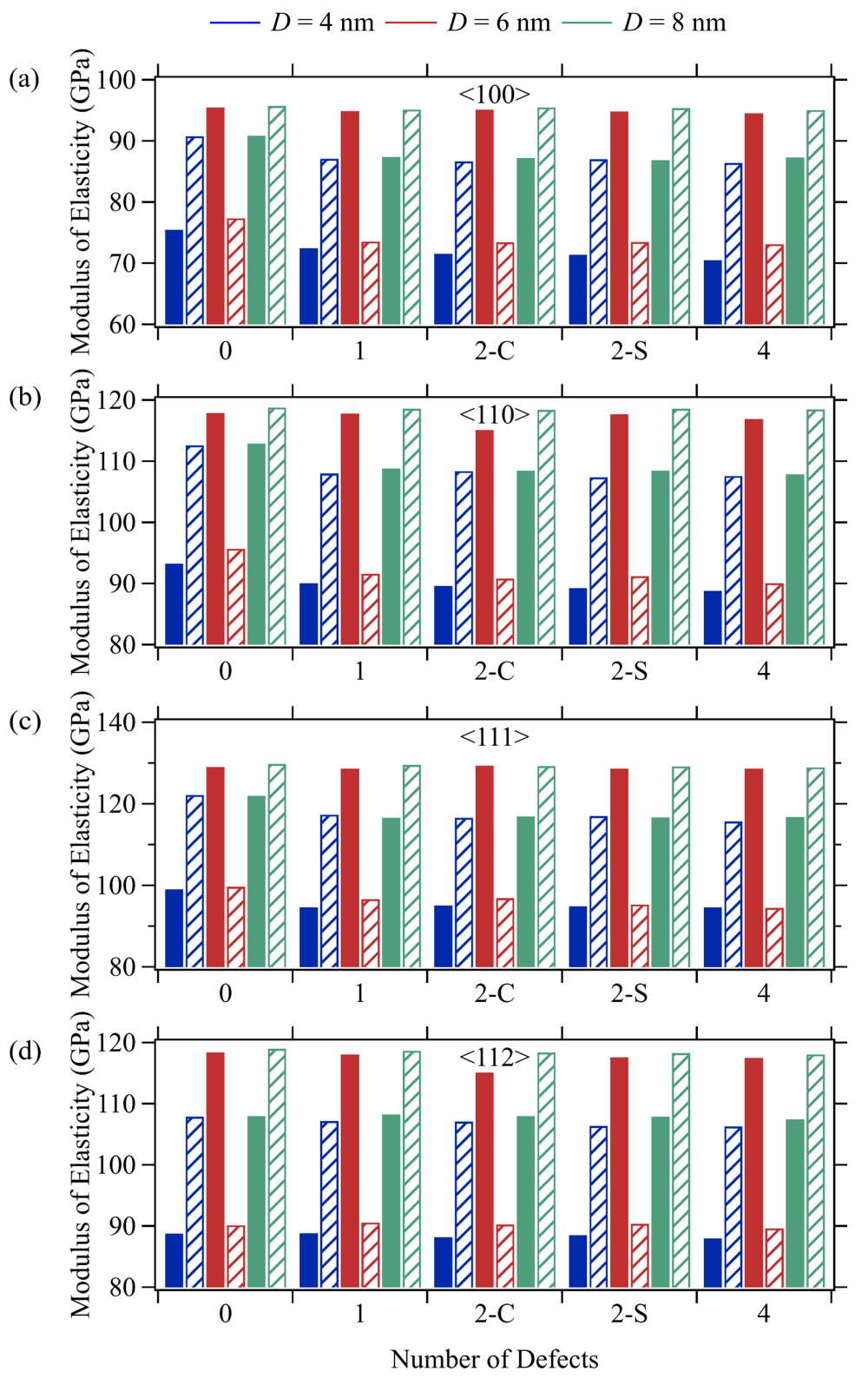}
	\caption{\label{fig:fig7} The modulus of elasticity as a function of defects distribution for aSiO$_2$-Si NWs oriented along (a) $<100>$, (b) $<110>$, (c) $<111>$, and (d) $<112>$ crystal orientations. The filled bars represent circular NWs, while the dashed bars correspond to square NWs.}
\end{figure}

\begin{figure}[ht]
	\centering
	\includegraphics[width=0.5\textwidth]{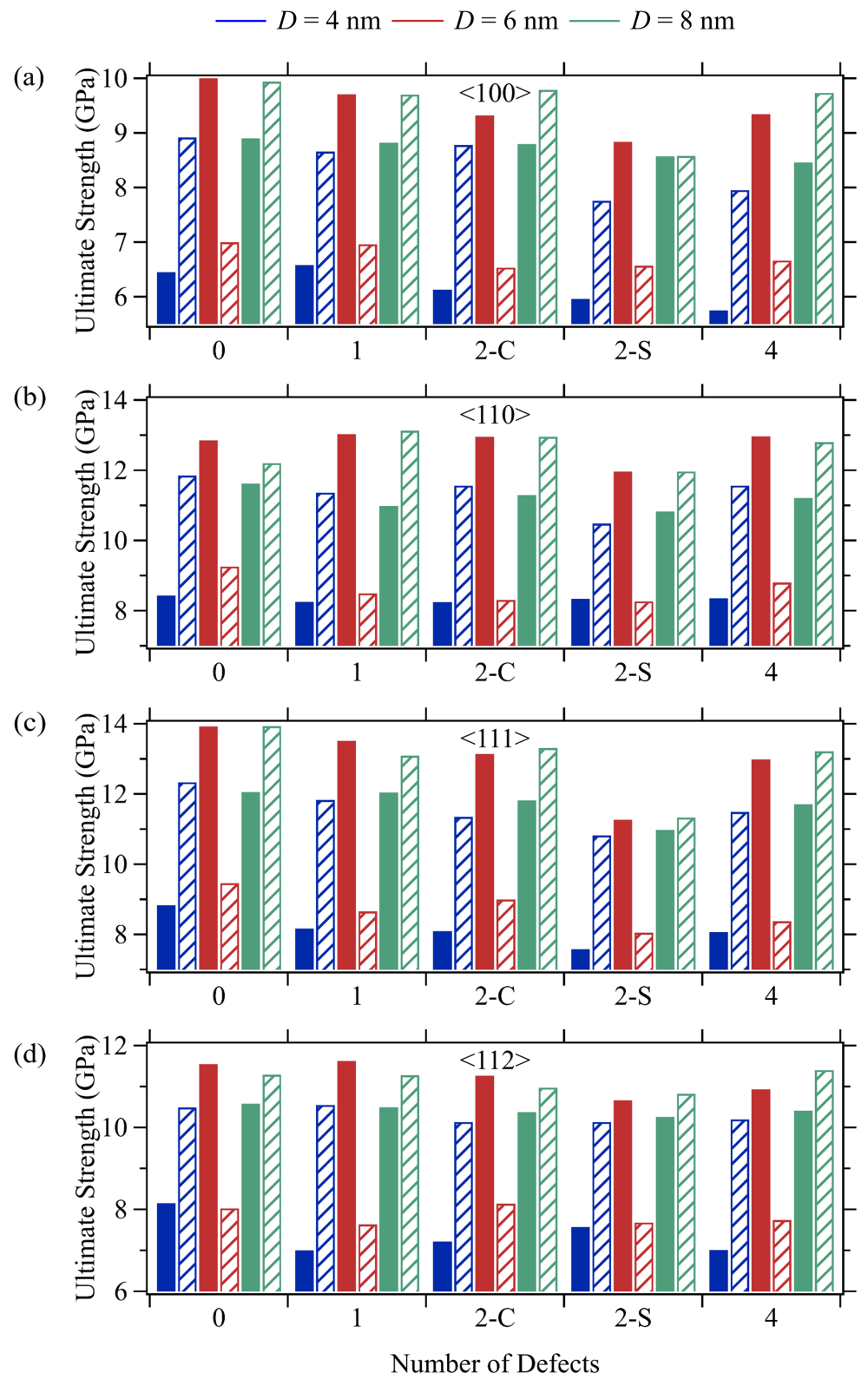}
	\caption{\label{fig:fig8} The ultimate strength as a function of defects distribution for aSiO$_2$-Si NWs oriented along (a) $<100>$, (b) $<110>$, (c) $<111>$, and (d) $<112>$ crystal orientations. The filled bars represent circular NWs, while the dashed bars correspond to square NWs.}
\end{figure}

In order to assess the impact of nano-voids on the tensile properties of aSiO$_2$-Si NWs, this study examines four distinct combinations placed along the NWs shown in Figure~\ref{fig:fig7} (d). Figure~\ref{fig:fig7} depicts the estimated modulus of elasticity for different nano-void combinations compared to the defect-free case in aSiO$_2$-Si NWs oriented along crystal orientations $<100>$, $<110>$, $<111>$, and $<112>$, as shown in Figure~\ref{fig:fig7} (a), (b), (c), and (d), respectively. A slight decrease in the modulus of elasticity is observed in aSiO$_2$-Si NWs as the number of voids increases from zero in the defect-free case to four nano-voids. This decrease becomes more negligible as $D$ increases, indicating that defects have a more significant impact at smaller critical dimensions. More precisely, in the case of aSiO$_2$-Si BWs with a $D$ of 4 nm, the modulus of elasticity experiences a reduction of up to 5 GPa with the inclusion of four nano-voids. The reduction is estimated to be 1-4 GPa for NWs with $D$ = 6 nm and 1-2 GPa for NWs with $D$ = 8 nm. Comparing the location of nano-voids near the surface (2-S) to centered ones (2-C), a slightly lower modulus of elasticity is observed when the nano-voids are closer to the surface. Figure~\ref{fig:fig8} illustrates the estimated ultimate strength for different nano-void combinations compared to the defect-free case in aSiO$_2$-Si NWs oriented along crystal orientations $<100>$, $<110>$, $<111>$, and $<112>$, as shown in Figure~\ref{fig:fig8} (a), (b), (c), and (d), respectively. A decreasing trend in the ultimate strength of aSiO$_2$-Si NWs, up to 2 GPa, is observed for different $D$ and crystal orientations. The ultimate strength exhibits a clear decrease as the number of nano-voids increases. Additionally, for aSiO$_2$-Si NWs with $D$ of 6 nm and 8 nm, the location of the nano-voids also affects the ultimate strength, with a noticeable decrease observed for nano-voids located near the surface (2-S) compared to the centered ones (2-C). This impact of void location is evident for different crystal orientations. A recent study investigating the effect of nano-voids on the modulus of elasticity in NWs reported reductions of up to 10 GPa, depending on the number and size of the nano-voids \cite{katakam2022tensile}.

\subsection{Strain Rate Effect}
\label{sec:3.5}

\begin{figure}[ht]
	\centering
	\includegraphics[width=0.5\textwidth]{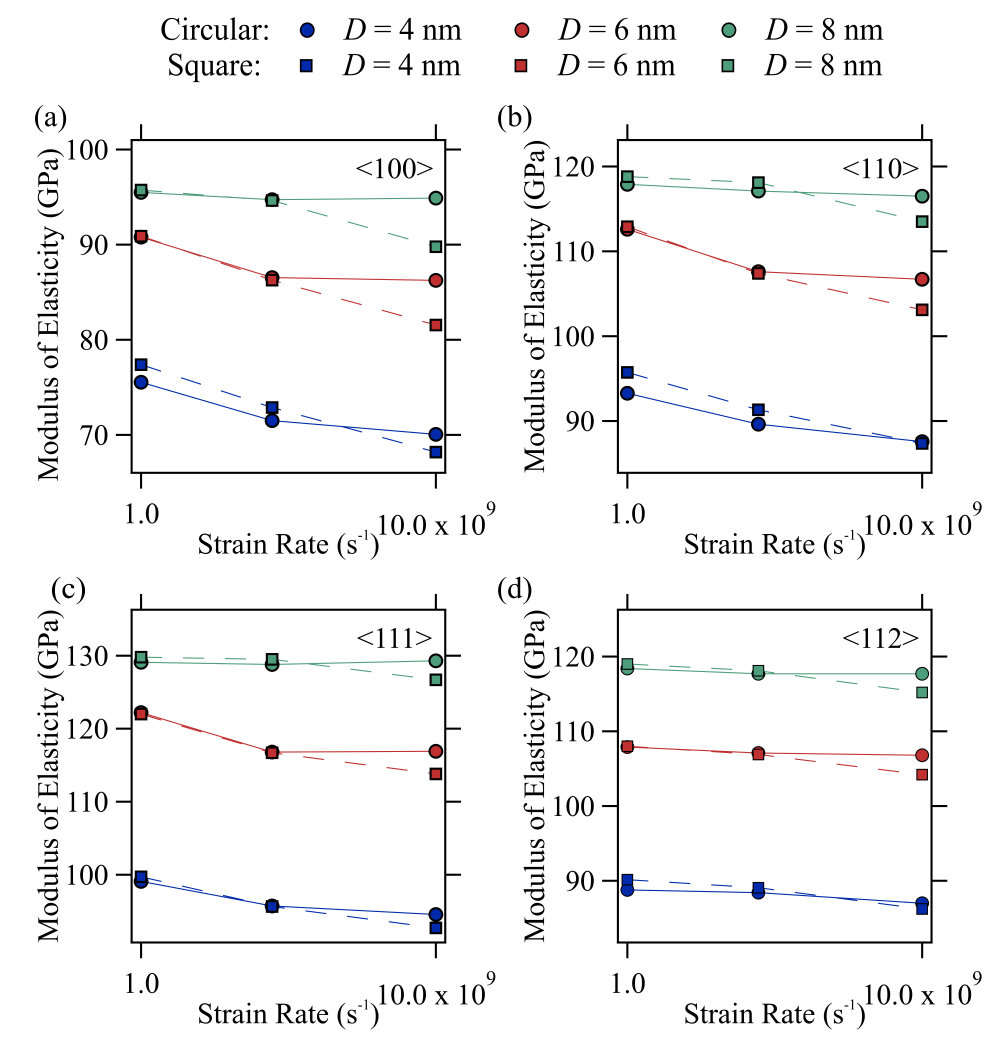}
	\caption{\label{fig:fig9} The modulus of elasticity as a function of strain rate for aSiO$_2$-Si NWs oriented along (a) $<100>$, (b) $<110>$, (c) $<111>$, and (d) $<112>$ crystal orientations.}
\end{figure}

\begin{figure}[ht]
	\centering
	\includegraphics[width=0.5\textwidth]{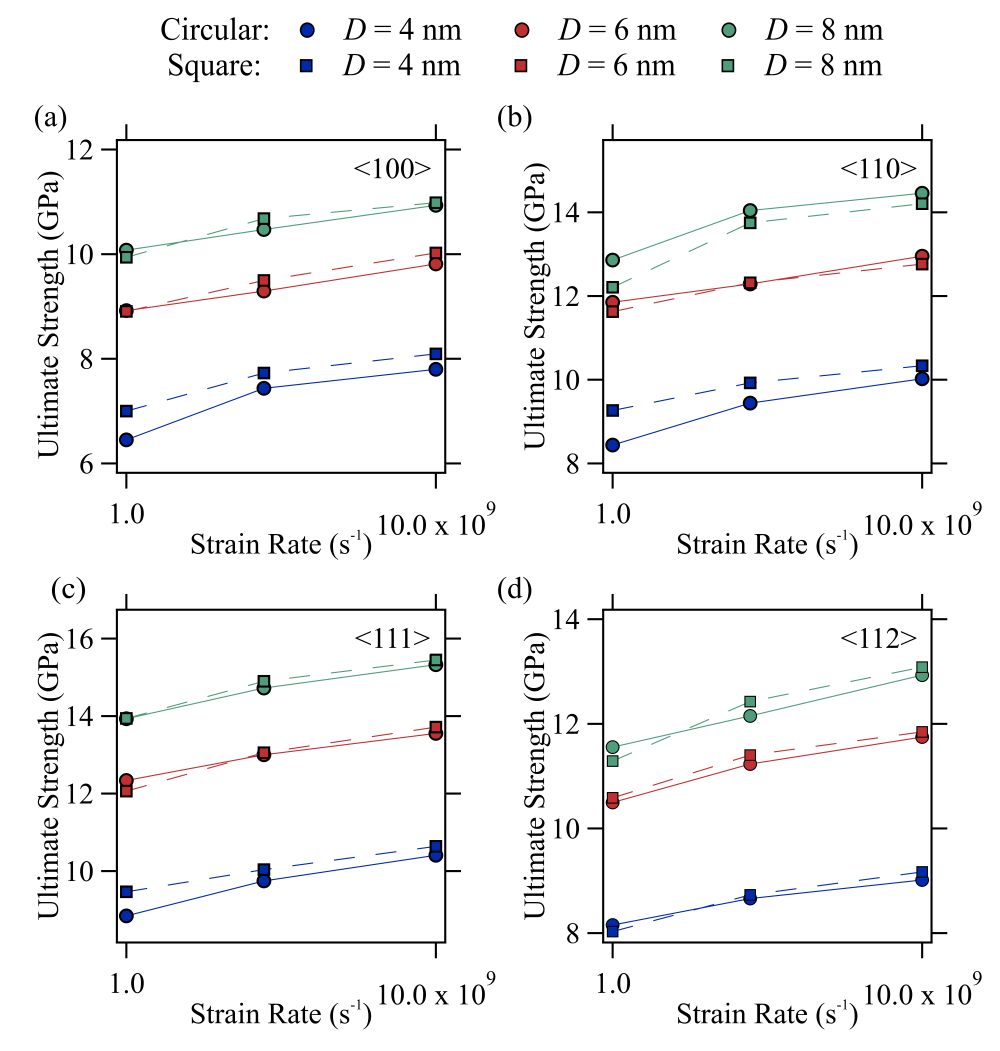}
	\caption{\label{fig:fig10} The ultimate strength as a function of strain rate for aSiO$_2$-Si NWs oriented along (a) $<100>$, (b) $<110>$, (c) $<111>$, and (d) $<112>$ crystal orientations.}
\end{figure}

The strain rate during tensile testing is another significant factor affecting the elastic properties and strength of aSiO$_2$-Si NWs. Figure~\ref{fig:fig9} illustrates the estimated modulus of elasticity at different strain rates for NWs oriented along crystal orientations $<100>$, $<110>$, $<111>$, and $<112>$, as shown in Figure~\ref{fig:fig9} (a), (b), (c), and (d), respectively. Tensile tests are conducted on aSiO$_2$-Si NWs using three strain rates: $\dot{\epsilon}$ =1 $\times$ 10$^{9}$ s$^{-1}$, $\dot{\epsilon}$ = 5 $\times$ 10$^{9}$ s$^{-1}$, and $\dot{\epsilon}$ = 1 $\times$ 10$^{10}$ s$^{-1}$. The results presented here, with minor variations due to the shape effect, indicate that the examined strain rates have an insignificant impact on the elastic properties of aSiO$_2$-Si NWs. Additionally, Figure~\ref{fig:fig10} displays the ultimate strength as a function of strain rate for NWs oriented along crystal orientations $<100>$, $<110>$, $<111>$, and $<112>$, as shown in Figure~\ref{fig:fig10} (a), (b), (c), and (d), respectively. The results reveal a slight increase in the ultimate strength of aSiO$_2$-Si NWs when subjected to higher strain rates. Specifically, the ultimate strength of aSiO$_2$-Si NWs tested at $\dot{\epsilon}$ = 1 $\times$ 10$^{10}$ s$^{-1}$ is found to be 1-3 GPa higher compared to NWs tested at $\dot{\epsilon}$ = 1 $\times$ 10$^{9}$ s$^{-1}$.

The investigation of strain rate effects on the tensile simulations of NWs is motivated by the need to assess the compatibility of interatomic potentials with the resulting elastic properties. An MD study employing the modified embedded-atom-method (MEAM) potentials reported negligible effects of strain rates spanning from $\dot{\epsilon}$ = 5 $\times$ 10$^{7}$ s$^{-1}$ to $\dot{\epsilon}$ = 5 $\times$ 10$^{8}$ s$^{-1}$ on the modulus of elasticity of unreconstructed Si NWs  \cite{xu2019molecular}. Another MD study on unreconstructed Si NWs, using the MEAM potential, found no significant influence of strain rates of $\dot{\epsilon}$ = 5 $\times$ 10$^{7}$ s$^{-1}$, $\dot{\epsilon}$ = 5 $\times$ 10$^{8}$ s$^{-1}$, and $\dot{\epsilon}$ = 5 $\times$ 10$^{9}$ s$^{-1}$ on the tensile behavior of NWs \cite{kang2010size}. Additionally, an MD study investigating the tensile response of unreconstructed Si NWs modeled with the SW potential determined that strain rates ranging from $\dot{\epsilon}$ = 2 $\times$ 10$^{8}$ s$^{-1}$ to $\dot{\epsilon}$ = 2 $\times$ 10$^{10}$ s$^{-1}$ had no significant effect on the estimated modulus of elasticity \cite{Jing2009, Jing2009b}. In the case of Si NWs with a native oxide surface state, a recent study utilizing the m-SW and Tersoff-Munetoh potentials found no significant impact of strain rates ranging from $\dot{\epsilon}$ = 1 $\times$ 10$^{8}$ s$^{-1}$ to $\dot{\epsilon}$ = 5 $\times$ 10$^{9}$ s$^{-1}$ on the elastic properties of the NWs \cite{pakzad2023role}.

Regarding the ultimate strength of Si NWs with a native oxide surface state, a recent study using the m-SW potential reported deviations of up to 15\% for strain rates ranging from $\dot{\epsilon}$ = 1 $\times$ 10$^{8}$ s$^{-1}$ to $\dot{\epsilon}$ = 5 $\times$ 10$^{9}$ s$^{-1}$ \cite{pakzad2023role}. Furthermore, the same study observed significant variations, up to 70\%, in ultimate strength due to strain rate effects when employing the Tersoff-Munetoh potential, which is known to overestimate the ultimate strength \cite{pakzad2023role}. Another study on unreconstructed Si NWs reported that a slower strain rate leads to a lower critical buckling load, allowing for sufficient local deformations in the Si NWs \cite{Jing2009, Jing2009b}. Similarly, a study on unreconstructed Si NWs found an approximate decrease of 1 GPa in ultimate strength as the strain rate decreases while providing more time for defects to nucleate at lower stresses \cite{kang2010size}. In the case of unreconstructed Si NWs modeled with modified MEAM potentials, no significant strain rate effects were observed for the ultimate strength when tested at $\dot{\epsilon}$ = 5 $\times$ 10$^{7}$ s$^{-1}$ and $\dot{\epsilon}$ = 5 $\times$ 10$^{8}$ s$^{-1}$ \cite{xu2019molecular}.

\section{Conclusion}
\label{sec:4}

In this study, the mechanical properties of aSiO$_2$-Si NWs are investigated by considering various factors such as crystallographic orientation, cross-sectional shape, critical dimensions, AR, temperature, and strain rate. Tensile tests are conducted on aSiO$_2$-Si NWs with circular and square cross-sections using different critical dimensions to understand their stress-strain behavior. The results revealed that the modulus of elasticity and ultimate strength of the NWs increased as the critical dimension increased while maintaining a constant AR and native oxide thickness. The crystallographic orientation and cross-sectional shape also exhibit significant effects on the mechanical properties of Si NWs. Furthermore, the AR shows a negligible effect on the modulus of elasticity for values above 12, while NWs with smaller ARs demonstrate higher modulus of elasticity. The elastic properties and strength of Si NWs are analyzed by investigating various combinations of nano-voids positioned along the NWs. The study also examines the influence of strain rate on the obtained results. Temperature dependence studies show decrease in the modulus of elasticity and ultimate strength with increasing temperature. The findings contribute to a better understanding of the mechanical behavior of aSiO$_2$-Si NWs, highlighting the importance of considering native oxide surface conditions and appropriate modeling parameters. This study lays the groundwork for further research and optimization of aSiO$_2$-Si NWs in various applications, providing insights for the design and development of high-performance Si NW-based devices.

\section*{ACKNOWLEDGMENT}

S.Z.P. and B.E.A. gratefully acknowledge the financial support by Tubitak under grant no. 120E347.

\bibliographystyle{IEEEtran}
\bibliography{SiNW_SZP}

\end{document}